# High-pressure Phase Transition of Olivine-type Mg$_2$GeO$_4$ to a Metastable Forsterite-III type Structure and their Equation of States.

R. Valli Divya[1], Gulshan Kumar[1], R. E. Cohen[2], Sally J. Tracy[2], Yue Meng[3], Stella Chariton[4], Vitali B. Prakapenka[4], and Rajkrishna Dutta[1, *]

[1]Department of Earth Sciences, IIT Gandhinagar, Gujarat 382355, India.
[2]Earth and Planets Laboratory, Carnegie Institution for Science, Washington DC 20015, USA.
[3]HPCAT, Advanced Photon Source, Argonne National Laboratory, Argonne, IL 60439, USA.
[4]Center for Advanced Radiation Sources, University of Chicago, Chicago, IL 60637, USA.

**Abstract**

Germanates are often used as structural analogs of planetary silicates. We have explored the high-pressure phase relations in Mg$_2$GeO$_4$ using diamond anvil cell experiments combined with synchrotron x-ray diffraction and computations based on density functional theory. Upon room temperature compression, forsterite-type Mg$_2$GeO$_4$ remains stable up to 30 GPa. At higher pressures, a phase transition to a forsterite-III type ($Cmc2_1$) structure was observed, which remained stable to the peak pressure of 105 GPa. Using a 3$^{rd}$ order Birch Murnaghan fit to the experimental data, we obtained $V_0$ = 305.1 (3) Å$^3$, $K_0$ = 124.6 (14) GPa and $K_0'$ = 3.86 (fixed) for forsterite- and $V_0$ = 263.5 (15) Å$^3$, $K_0$ = 175 (7) GPa and $K_0'$ = 4.2 (fixed) for the forsterite-III type phase. The forsterite-III type structure was found to be metastable when compared to the stable assemblage of perovskite/post-perovskite + MgO, as observed during laser-heating experiments. Understanding the phase relations and physical properties of metastable phases is crucial for studying the mineralogy of impact sites, understanding metastable wedges in subducting slabs and interpreting the results of shock compression experiments.

## 1. Introduction

(Mg,Fe)$_2$SiO$_4$ olivine is the most abundant mineral in the Earth's upper mantle. The major seismic discontinuities (410, 520 and 660 km) in the upper mantle and transition zone can be attributed to pressure induced phase transitions in Mg-rich olivine to β-olivine (wadsleyite), γ-olivine (ringwoodite) and (Mg, Fe)SiO$_3$ perovskite (bridgmanite, Pv) + (Mg, Fe)O magnesiowüstite (Ringwood 1991). The D" layer, located in the lowermost ~250 km of the mantle is characterized by a transition from bridgmanite to post-perovskite (pPv; 125 GPa and 2500 K; Murakami et al. 2004; Oganov and Ono 2004; Tsuchiya et al. 2004). Post-perovskite (Mg, Fe)SiO$_3$ is expected to be the highest-pressure silicate phase in the Earth. However, in the case of terrestrial Super-Earth planets, where the pressure-temperature conditions at the core-mantle boundary can be substantially higher (e.g., > 1600 GPa and ~6500 K for a planet with a mass equivalent to that of 10 Earths; van den Berg et al. 2019), additional transitions are possible. At ~500 GPa, pPv + MgO is expected to recombine into a tetragonal $I\bar{4}2d$ or cubic $I\bar{4}3d$ Mg$_2$SiO$_4$ phase (Umemoto et al., 2017; Dutta et al., 2023), followed by a dissociation into the binary oxides at ~3000 GPa. However, all the post-pPv transitions have only been computationally predicted and not observed experimentally because of the extreme pressure-temperature conditions, which are beyond the limits of conventional experimental techniques. As an alternative, silicate analogs like germanates (Ringwood & Seabrook, 1963; Umemoto & Wentzcovitch, 2019; Dutta et al., 2018; Dutta et al., 2022) and fluorides (Grocholski et al., 2010; Dutta et al., 2019) can be used in high-pressure experiments as they undergo similar phase transitions, but at significantly lower pressures, e.g. the Pv-pPv phase transition, which occurs at 125 GPa in MgSiO$_3$ is observed at 65 GPa in the germanate (Hirose et al. 2005). Additionally, the $I\bar{4}2d$/ $I\bar{4}3d$ phase in Mg$_2$GeO$_4$ has been reported at pressures > 170 GPa from experiments (Dutta et al., 2022) and computational studies (Umemoto and Wentzcovitch 2019, 2021) in comparison to the theoretical prediction of 0.5 TPa in the silicate (Umemoto et al. 2017).

There is considerable interest in understanding the 300 K compression behavior of both the silicate and germanate olivine as well. Knowledge of the metastable transitions in olivine can help in understanding mineral phases formed at impact sites (Van de Moortèle et al. 2007). It is potentially useful in inferring phase transitions in laboratory shock experiments, where the short time scale may prevent formation of stable assemblages (Kim et al. 2021). In $Mg_2SiO_4$, existing studies have reported pressure induced amorphization (Guyot and Reynard 1992; Andrault et al. 1995), change in compression mechanism (Rouquette et al. 2008) or recently a transition to forsterite-II and forsterite-III (Finkelstein et al., 2014; referred to as Fo-II and Fo-III after this) structures. In $Mg_2GeO_4$, the stable phase at ambient pressure and low temperatures is the spinel structure (Ross and Navrotsky 1987). The high-temperature phase, olivine reverts to the spinel phase at 1083 K (Dachille and Roy 1960) and persists on quenching to ambient temperature. On compressing olivine at room-temperatures, it has been reported to stay stable up to 13 GPa, after which new diffraction peaks were observed (Petit et al. 1996) and could not be resolved. Pressure-induced amorphization has been reported above 22-25 GPa (Petit et al. 1996; Nagai et al. 1994). High-pressure Raman spectroscopic studies have observed appearance of new modes at ~11 GPa, followed by a sharp decrease in its intensity at ~25 GPa (Reynard et al. 1994). In this work we aim to resolve the post-olivine structure(s) under compression by studying forsterite-type $Mg_2GeO_4$ to 105 GPa at both room and high-temperature using laser-heated diamond anvil cells (LH-DAC) and density functional theory (DFT) based computations.

## 2. Methodology

### A. Experimental details

The starting material, $Mg_2GeO_4$ olivine was synthesized by heating high-purity MgO and $GeO_2$ to 1473 K for 5 days (Ross and Navrotsky 1987; Dutta et al. 2022) and confirmed using Raman spectroscopy and X-ray diffraction. The synthesized sample was ground with 10 wt% gold, which acts as the laser absorber and pressure marker during the high-pressure experiments.

The samples were compressed using symmetric diamond anvil cells with 300 μm – 150 μm diameter culets. Rhenium gaskets were drilled to form the sample chamber. The samples were loaded inside the sample cavities (200 – 80 μm) and gas loaded with Ne to provide a quasi-hydrostatic environment. *In situ* X-ray diffraction (XRD) was carried out at sectors 13-ID-D and 16-ID-B of the Advanced Photon Source using a monochromatic beam with wavelengths of 0.2952 Å and 0.4066 Å respectively. The two-dimensional X-ray images were radially integrated to the one dimensional patterns using DIOPTAS (Prescher and Prakapenka 2015). Double sided laser heating was used to produce the high temperatures. Temperatures were increased in small steps of ~100 K and measured using spectroradiometry (Jephcoat and Besedin 1996; Shen et al. 2001). The (111) Au peak was used to calculate the pressures (Fei et al. 2007) using the Birch Murnaghan equation of state (EOS). The lattice parameters were calculated using least squares refinement of the peak positions (Holland and Redfern 1997) fitted to Voigt line shapes or whole profile Le Bail refinement as implemented in the GSAS-II package (Toby and Von Dreele 2013). The background was fitted with a $6^{th}$ order Chebyschev polynomial. The unit cell dimensions, instrumental and sample broadening parameters were initially refined separately and then together.

B. **Computational details**

All computations were performed using the plane wave implementation of density functional theory through the Quantum Espresso package (Giannozzi et al. 2009). The generalized gradient approximation (GGA-PBE) (Perdew et al. 1996) was used to treat the exchange and correlation functional. We have used a plane wave basis set with a cutoff of 40 Ry and a Monkhorst-Pack (Monkhorst and Pack 1976) *k*-point grid of 6x6x6 for all the considered structures. Ultrasoft pseudopotentials (Vanderbilt 1990) were used to describe the electron-ion interactions. The geometry optimizations were carried out using the BFGS algorithm (Broyden 1970) by relaxing the lattice parameters and atomic positions at each pressure step. The structural relaxations were considered complete when the forces on atoms were less than $1\times10^{-4}$ Ry/Bohr and total energies were converged to $1\times10^{-6}$ Ry.

## 3. Results

In three separate experimental runs, the germanate olivine samples were compressed to peak pressures of 26 GPa, 54 GPa and 105 GPa at room-temperature (Fig. 1). The diffraction patterns up to 30 GPa can be indexed using the ambient-pressure olivine structure, suggesting a metastable persistence. As an example, table S1 of the supplementary material shows the observed and calculated *d*-spacings for forsterite $Mg_2GeO_4$ at 14.6 GPa. The difference between the two values is < 0.002 Å, suggesting a good fit of the olivine structure to observed pattern. This is also reflected in the whole profile Le Bail refinement of the measured pattern at 26 GPa (Fig. 2). In contrast to previous studies (Nagai et al. 1994; Petit et al. 1996), we did not find any evidence for amorphization. The lattice parameters of $Mg_2GeO_4$ olivine at 26 GPa are $a$ = 4.7573 Å, $b$ = 9.6574 Å and $c$ = 5.7064 Å. Figure 3 and table S2 of the supplementary material shows the change in the unit cell dimensions as a function of pressure. Although our work extends to higher pressures, it is in fair agreement with existing experimental studies, especially at lower pressures. At higher pressure, the discrepancy possibly arises from the non-hydrostatic conditions (Klotz et al. 2009) inside the DAC in the previous work. The linear compressibilities ($\times 10^{-3}$ GPa$^{-1}$) of the axes for the experimental (theoretical) are $\beta_a$ = 1.21 (1.15), $\beta_b$ = 2.29 (2.35), $\beta_c$ = 1.98 (1.96). Despite the GGA's tendency to overestimate the lattice parameters, the remarkable concurrence of experimental and computed linear compressibilities emphasizes their strong agreement. The order of the axial compressibilities i.e. $\beta_b > \beta_c > \beta_a$ also agree with that of $Mg_2SiO_4$ forsterite (Zhang 1998; Finkelstein et al. 2014).

Upon further compression to 40 GPa (Fig. 1), new diffraction peaks were observed, which were retained up to the peak pressure of 105 GPa. To understand the structure of the new phase, we computed the enthalpies (Fig. 4) of spinel and several post-spinel $Mg_2GeO_4$ phases. The structures were derived from related systems e.g. Fo-II and Fo-III (Finkelstein et al., 2014), Fo-IV (Bouibes & Zaoui, 2020), $I\bar{4}2d$ (Dutta et al., 2022), Pv + MgO (Leinenweber et al. 1994), pPv + MgO (Hirose et al., 2005), *Pmma* $CaTi_2O_4$-type (Yamanaka et al., 2013), $CaFe_2O_4$-type (Decker and Kasper 1957) and the $Ca_2IrO_4$-type (Babel et al. 1966) structures. It can be seen the Pv + MgO assemblage becomes more stable (lower enthalpy) with respect to the olivine-type $Mg_2GeO_4$ structure at ~12 GPa, which then transforms into the pPv structure at ~50 GPa. Taking into account the tendency of the GGA-PBE functional to underestimate transition pressures, these results can be viewed as reasonably consistent with experimental findings (Liu 1977; Hirose et al. 2005). The XRD patterns at P > 40 GPa are not consistent with any of these phases, suggesting the presence of a metastable phase. This can be attributed to the experimental conditions being at room temperature, which creates a kinetic barrier that prevents the transition to the more stable assemblage. Besides Pv and pPv, the candidate phases with low enthalpies are the Fo-II type, Fo-III type and $CaTi_2O_4$-type $Mg_2GeO_4$ structures. Figure 5 compares the observed XRD pattern at 61 GPa with the simulated diffraction pattern of these three phases. In agreement with a previous theoretical study (Bouibes and Zaoui 2020) on $Mg_2SiO_4$, the triclinic Fo-II structure (Finkelstein et al. 2014) was neither energetically favored computationally, nor did it match the XRD data. The closest match to the observed patterns were the ordered *Pmma* $CaTi_2O_4$-type phase and the Fo-III phase. Although the simulated patterns for the two are similar, the Fo-III structure (CIF on deposit, optimized DFT structure at 60 GPa) is a better match (fewer peaks) and comparatively lower enthalpy.

Post-spinel (e.g. $CaMn_2O_4$-, $CaFe_2O_4$- and $CaTi_2O_4$- type) structures (Yamanaka et al. 2008) generally feature chains of octahedra that share edges and corners, forming channels that align parallel to the *c*-axis. The Fo-III structure (Fig. 6) is analogous to an inverse spinel structure. It is related to the non-centrosymmetric variant of the *Cmcm* $CaTi_2O_4$ post-spinel structure (Yamanaka et al. 2013) in which half of Mg atoms are situated in the larger trigonal prismatic site (Mg2), while the other half occupy the octahedral (Mg1) site (Finkelstein et al. 2014).This is substantially different from the olivine structure, where both the Mg1 and Mg2 sites are octahedral with one being more distorted than the other. The Fo-III structure also marks an increase in the Ge-coordination from 4 (as in olivine) to 6, providing a pathway to the stable six-coordinated pv and pPv structures. The structural parameters of Fo- and Fo-III type $Mg_2GeO_4$ have been shown in Table 1. Figure 7 shows a Le Bail refinement of the measured diffraction pattern of $Mg_2GeO_4$ at 74 GPa. The difference between the calculated and observed *d*-spacings were less than < 0.006 Å (Table S3 of the supplementary material, 68 GPa), again suggesting a good fit of the measured diffraction patterns with the Fo-III structure. Figure 8 and table S4 of the supplementary material shows the variation in lattice parameters of Fo-III with increasing pressure. The experimental *a*, *b* and *c* parameters are found to decrease by 2.9%, 2.7% and 2.7% respectively over the pressure range (40.4 GPa – 73.8 GPa) considered. The theoretical axial parameters decrease by 3.3%, 3.7% and 3.3% respectively between 40 and 80 GPa, indicating a good agreement with the experiments. No further transitions were observed up to the peak pressure of 105 GPa at room-temperature.

The pressure-volume data for both the Fo- and Fo-III type $Mg_2GeO_4$ phases (Fig. 9) were fitted to an isothermal 3$^{rd}$ order Birch Murnaghan (BM) equation of state. Table 2 presents the EOS parameters for these phases and includes a comparison with the existing studies on the same structures in $Mg_2GeO_4$ (Weidner and Hamaya 1983; Nagai et al. 1994; Petit et al. 1996) and $Mg_2SiO_4$ (Andrault et al. 1995; Downs et al. 1996; Zhang 1998; Finkelstein et al. 2014; Zhang et al. 2019; Bouibes and Zaoui 2020). For the germanate olivine, the EOS parameters for the computed data are $V_0$ = 316.8 (3) Å$^3$, $K_0$ = 112.2 (13) GPa and $K'_0$ = 3.86 (5), where $V_0$, $K_0$ and $K'_0$ are the unit cell volume, bulk modulus, and its pressure derivative at ambient pressure respectively. In case of the experimentally obtained values, the $K'_0$ was fixed to the theoretical value of 3.86. This yielded $V_0$ = 305.1 (3) and $K_0$ = 124.6 (14) GPa. This is in excellent agreement with existing ultrasonic ($K_0$ = 120 GPa; Soga, 1971) and Brillouin spectroscopic measurements ($K_0$ = 120 GPa; Weidner & Hamaya, 1983). However, the obtained bulk modulus is significantly less than that obtained from previous DAC studies ($K_0$ = 166 (15) at fixed $K'_0$ = 4; Petit et al., 1996). The difference probably arises from the limited pressure coverage in the previous work along with the use of silicone oil, which is known to provide limited hydrostaticity at high-pressures (Klotz et al. 2009). The EOS parameters are also in good agreement with the silicate olivine ($K_0$ = 130.0 (9) GPa and $K'_0$ = 4.12 (7); Finkelstein et al., 2014). The transition from forsterite- to Fo-III type $Mg_2GeO_4$ is expected to have a substantial volume change of 9.53% at 35 GPa, which is in excellent agreement with its silicate counterpart (8.3% at 58 GPa). In case of Fo-III $Mg_2GeO_4$, the EOS parameters for the theoretical data are: $V_0$ = 271. 8 (9) Å$^3$, $K_0$ = 162.9 (5) GPa and $K'_0$ = 4.19 (1). The fit to the experimental data yielded $V_0$ = 263.5 (15) Å$^3$, $K_0$ = 175 (7) GPa, with $K'_0$ fixed to the computed value (4.19). These values are in fair agreement with the theoretical EOS parameters for $Mg_2SiO_4$ ($V_0$ = 247.4517 Å$^3$, $K_0$ = 197.12 GPa and $K_0'$ = 3.4, (Bouibes and Zaoui 2020).

On laser-heating the $Mg_2GeO_4$ sample to 2331 ± 148 K for 2-5 minutes at 26 GPa, we observed new diffraction peaks that could not be explained using forsterite-, spinel- or forsterite-III type $Mg_2GeO_4$ structures. The XRD peaks were instead consistent with an assemblage of Pv-$MgGeO_3$ + B1-MgO. This is in good agreement with our computations which predict a transition from the olivine-type structure to Pv-$MgGeO_3$ + MgO at 12 GPa and existing experimental studies with a olivine-type starting material (26 GPa; Liu 1977) as well as a $MgGeO_3$ pyroxene starting material (25 GPa; Runge et al. 2006). The sample was further compressed to 54 GPa at room temperature, followed by heating a fresh spot to a peak temperature of 2463 ± 112 K in small steps of 200 K. The observed diffraction pattern could still be indexed using $MgGeO_3$-Pv + B1-MgO. Figure 10 shows a Le Bail refinement of the XRD pattern at 65 GPa. The lattice obtained from the refinement ($a$ = 4.584 Å, $b$ = 4.858 Å, $c$ = 6.727 Å) are in excellent agreement with previous studies ($a$ = 4.587 Å, $b$ = 4.860 Å, $c$ = 6.721 Å at 65.7 GPa, Runge et al. 2006). In the experiment where a fresh sample was compressed to 105 GPa at room temperature and subsequently heated to 2280 ± 46 K, the diffraction pattern could be explained using a mixture of $CaIrO_3$-type post-perovskite $MgGeO_3$ + B1-MgO (Figure 11). This is consistent with the reported Pv to pPv transition pressure of 63 GPa with a orthoenstatite starting material (Hirose et al. 2005). The lattice parameters obtained from the Le Bail refinement at 110 GPa, 2300 K are ($a$ = 2.567 Å, $b$ = 8.301 Å, $c$ = 6.351 Å) are in fair agreement with existing experimental work ($a$ = 2.575 Å, $b$ = 8.324 Å, $c$ = 6.349 Å at 107 GPa and 300 K, (Kubo et al. 2006).

## 4. Discussion and Implications

Knowledge of metastable phases are important for understanding the mineralogy of planetary impact sites and meteorites e.g. Martian meteorites NWA 2737 and NWA 1950 (Van de Moortèle et al. 2007). The ultrafast timescales of dynamic compression experiments are often not enough to stabilize the equilibrium stable structures, leading to formation of metastable phases. The metastable olivine wedge hypothesis (Soga 1971; Däßler and Yuen 1996) has commonly been used to explain stagnation of subducting slabs and origin of deep-focus earthquakes. The P, T conditions in the cold subducting slabs may also stabilize metastable phases like Forsterite-III and thereby contribute to the high seismic velocities observed near the 660 km discontinuity (Zhang et al. 2019).

Recent laser-based shock compression experiments (Kim et al. 2021) on forsterite $Mg_2SiO_4$ have shown the presence of a metastable Fo-III phase instead of the stable assemblage i.e. bridgmanite + MgO at pressures > 33 GPa. $Mg_2GeO_4$ olivine is a widely used analog for forsterite $Mg_2SiO_4$ and is expected to show similar phase transitions, but at lower pressures. The high-pressure data on the germanate olivine is limited to pressures < 35 GPa and suggest a pressure induced amorphization under compression at room temperature (Nagai et al. 1994; Petit et al. 1996). Using synchrotron X-ray diffraction measurements and density functional computations, we have shown that $Mg_2GeO_4$ olivine persists metastably up to 30 GPa. It then undergoes a pressure induced phase transition to a metastable forsterite-III structure. Forsterite-III stays stable up to the peak pressure of 105 GPa (at 300 K), with no evidences of the forsterite-II (Finkelstein et al. 2014) phase seen in the silicate or pressure induced amorphization. We have also obtained equation of state parameters of both the forsterite and forsterite-III phases. Although, our bulk modulus value for forsterite ($K_0$ = 124.6 GPa) is lower than previous high-pressure studies (e.g. $K_0$ = 166 GPa, Petit et al. 1996), it is in excellent agreement with ultrasonic ($K_0$ = 120 GPa; Soga, 1971) and Brillouin spectroscopic measurements ($K_0$ = 120 GPa; Weidner & Hamaya, 1983). The enhanced quality of our calculated EOS parameters can be attributed to the utilization of a wider data range and the incorporation of a more hydrostatic pressure medium (Ne). To the best of our knowledge, there is no available data for Fo-III.

The Fo-III phase has now been reported in laser (~10 ns time scale, (Kim et al. 2021)) and gas gun (~100s of ns, (Newman et al. 2018)) based shock compression studies as well as static compression experiments in both silicates (Finkelstein et al. 2014) and germanates (this study). This suggests it may be an important transition to pathway to the stable higher-coordination structures at higher temperatures. On laser-heating at 26 and 54 GPa, a partial dissociation into bridgmanite $MgGeO_3$ + B1-MgO was observed. At 105 GPa, post-perovskite $MgGeO_3$ was observed instead of bridgmanite. The presence of both the perovskite and post-perovskite structures at high pressures and temperatures in $Mg_2GeO_4$ makes it an excellent low-pressure analog of $Mg_2SiO_4$.

## 5. Acknowledgements

We would like to thank F. Miozzi and J. Yang for help with experiments. We acknowledge the support of GeoSoilEnviroCARS (Sector 13), which is supported by the NSF - Earth Sciences (EAR-1128799), and the U.S. Department of Energy (DOE), Geosciences (DE-FG02-94ER14466). Portions of this work were performed at HPCAT (Sector 16), Advanced Photon Source (APS), Argonne National Laboratory. HPCAT operations are supported by DOE-NNSA's Office of Experimental Sciences. This research used resources of the Advanced Photon Source; a DOE Office of Science User Facility operated by Argonne National Laboratory under Contract No. DE-AC02-06CH11357. RD is grateful to SERB-Department of Science and Technology, India for financial support. RVD thanks CSIR-India for providing the research fellowship. RD and REC gratefully acknowledge the Gauss Centre for Supercomputing e.V. (www.gausscentre.eu) for funding this project by providing computing time on the GCS Supercomputer SuperMUC-NG at Leibniz Supercomputing Centre (LRZ, www.lrz.de).

**Figure Captions**

Figure 1. Select X-ray diffraction patterns of $Mg_2GeO_4$ under compression at room-temperature. Fo, Fo-III, Au, Ne and Re indicate the peaks from forsterite, forsterite-III, gold, neon, and rhenium respectively.

Figure 2. Le Bail refinement of the X-ray diffraction pattern of $Mg_2GeO_4$ at 26 GPa and 300 K. Black crosses show the observed spectrum. Red, green, and blue lines indicate the calculated spectrum, background, and difference between observed and fitted spectra respectively. The colored bars at the bottom show the different phases.

Figure 3. Change in lattice parameters of forsterite-type $Mg_2GeO_4$ with pressure. The solid data points represent this study (red: experiments, blue: DFT-PBE), while the open symbols show existing experimental studies (yellow Petit et al., 1996, purple: Nagai et al., 1994). The lattice parameters of $Mg_2SiO_4$ (Finkelstein et al., 2014) have also been shown for comparison (open green symbols).

Figure 4. Enthalpy difference of the different phases of $Mg_2GeO_4$ with respect to the Forsterite-III type phase at 0 K.

Figure 5. Comparison of the observed XRD pattern at 61 GPa with the simulated patterns of the computed Fo-II, Fo-III and *Pmma* $CaTi_2O_4$-type $Mg_2GeO_4$ structures at 60 GPa.

Figure 6. Crystal structure of forsterite- and forsterite-III type $Mg_2GeO_4$. Mg1 and Mg2 indicate the two inequivalent magnesium sites.

Figure 7. Le Bail refinement of XRD pattern of $Mg_2GeO_4$ at 74 GPa and 300 K. Colors have the same meaning as figure 2.

Figure 8. Lattice parameters of Fo-III type $Mg_2GeO_4$ versus pressure. Solid orange and blue indicate our experimental and theoretical data respectively. Open green symbols show the $Mg_2SiO_4$ data (Finkelstein et al., 2014).

Figure 9. Variation in unit cell volume as a function of pressure. Solid circles (red: Forsterite, orange: Fo-III) and triangles (dark blue: Fo, light blue: FoIII) represent experimental and theoretical data from this study. Solid lines are $3^{rd}$ order BM fits to the data. The two shades of green show the silicate data for Fo and Fo-III respectively (Finkelstein et al., 2014). Other colors have the same meaning as the figure 3.

Figure 10. Le Bail refinement of the X-ray diffraction pattern of $Mg_2GeO_4$ after laser heating to 2460 K and then quenching to room temperature at 65 GPa. Colors have the same meaning as figure 2.

Figure 11. Le Bail refinement of the diffraction pattern of $Mg_2GeO_4$ at 110 GPa and 2280 K. Colors have the same meaning as figure 2.

Figure 1. Select X-ray diffraction patterns of Mg$_2$GeO$_4$ under compression at room-temperature. Fo, Fo-III, Au, Ne and Re indicate the peaks from forsterite, forsterite-III, gold, neon, and rhenium respectively.

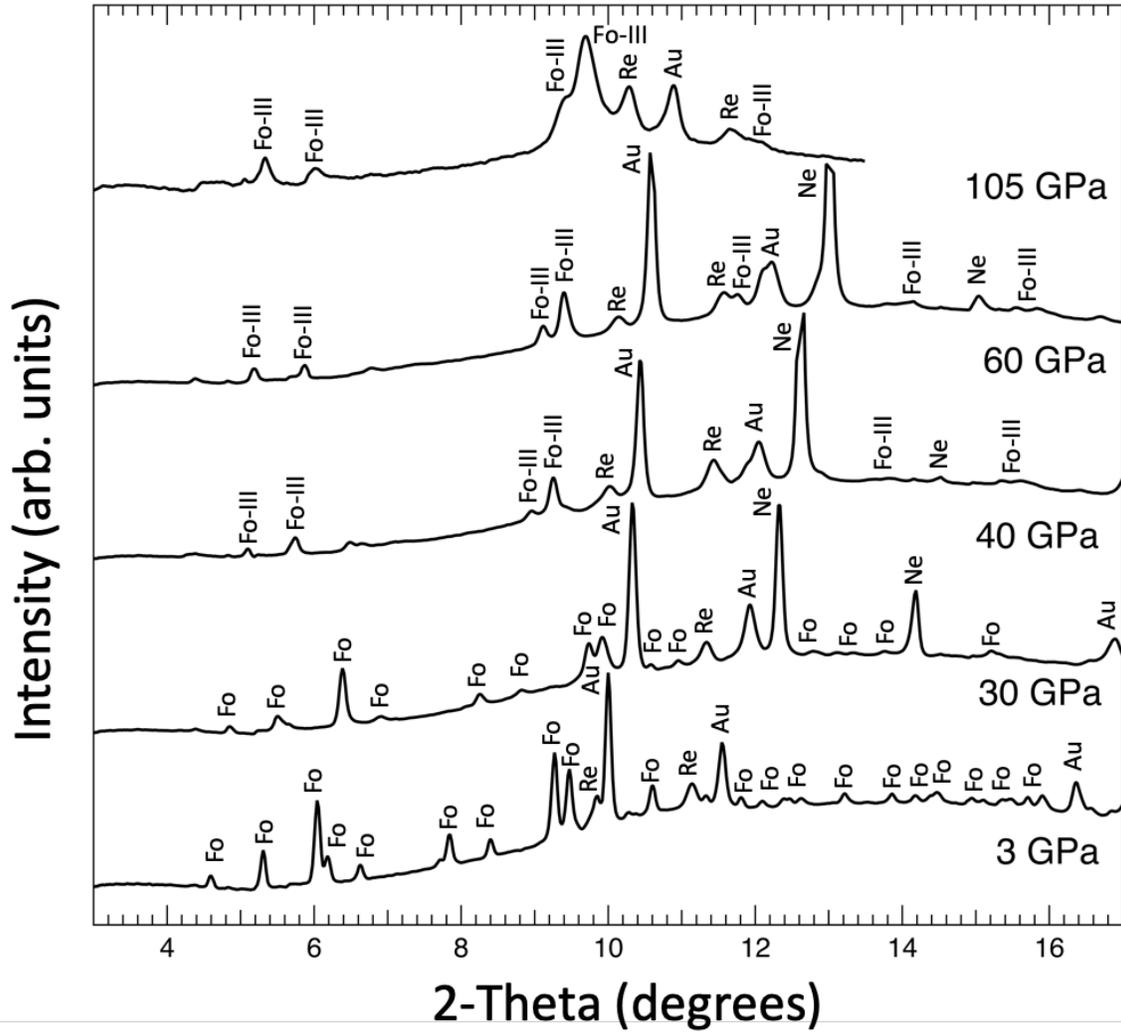

Figure 2. Le Bail refinement of the X-ray diffraction pattern of $Mg_2GeO_4$ at 26 GPa and 300 K. Black crosses show the observed spectrum. Red, green, and blue lines indicate the calculated spectrum, background, and difference between observed and fitted spectra respectively. The colored bars at the bottom show the different phases.

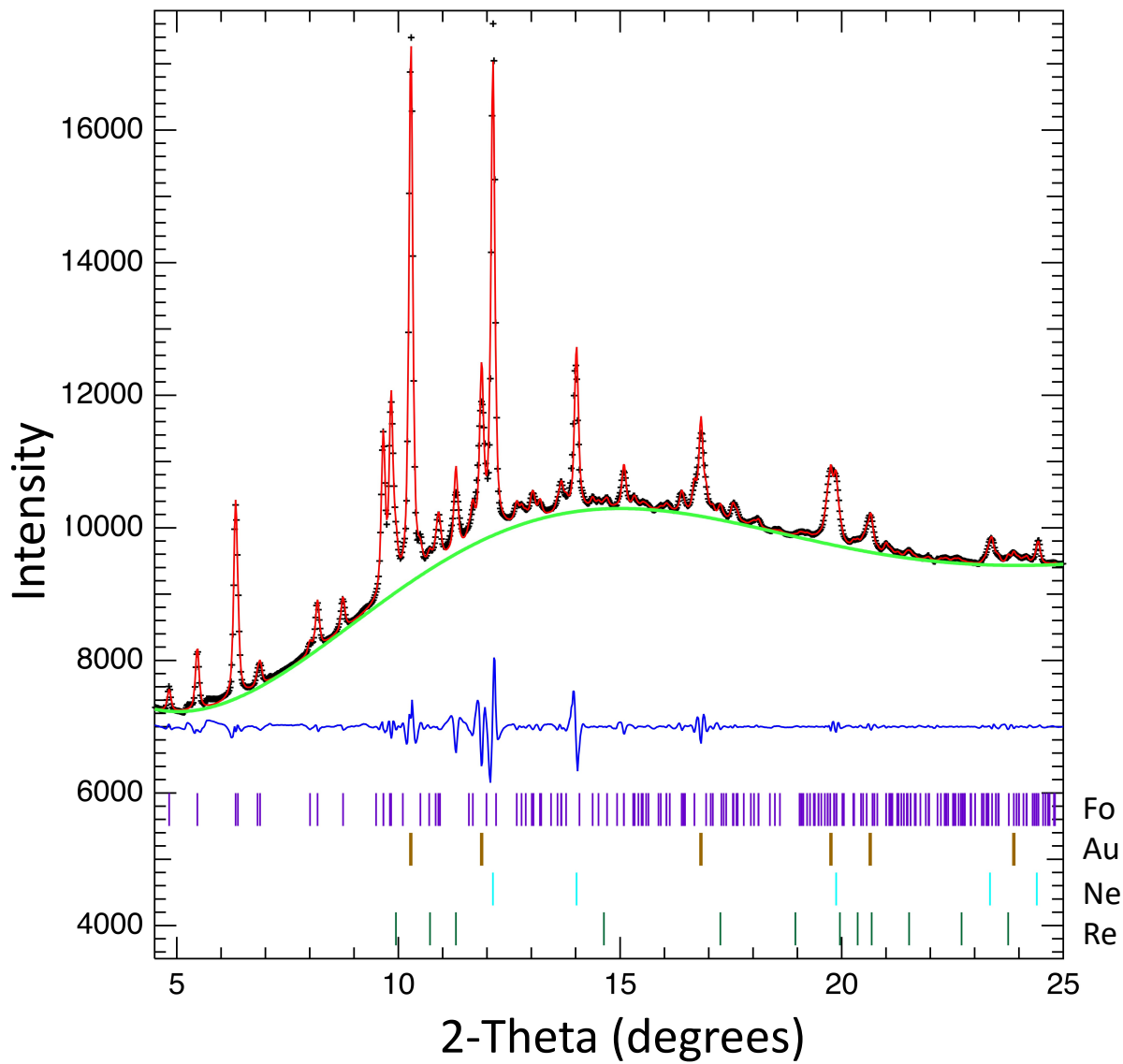

Figure 3. Change in lattice parameters of forsterite-type $Mg_2GeO_4$ with pressure. The solid data points represent this study (red: experiments, blue: DFT-PBE), while the open symbols show existing experimental studies (yellow Petit et al., 1996, purple: Nagai et al., 1994). The lattice parameters of $Mg_2SiO_4$ (Finkelstein et al., 2014) have also been shown for comparison (open green symbols).

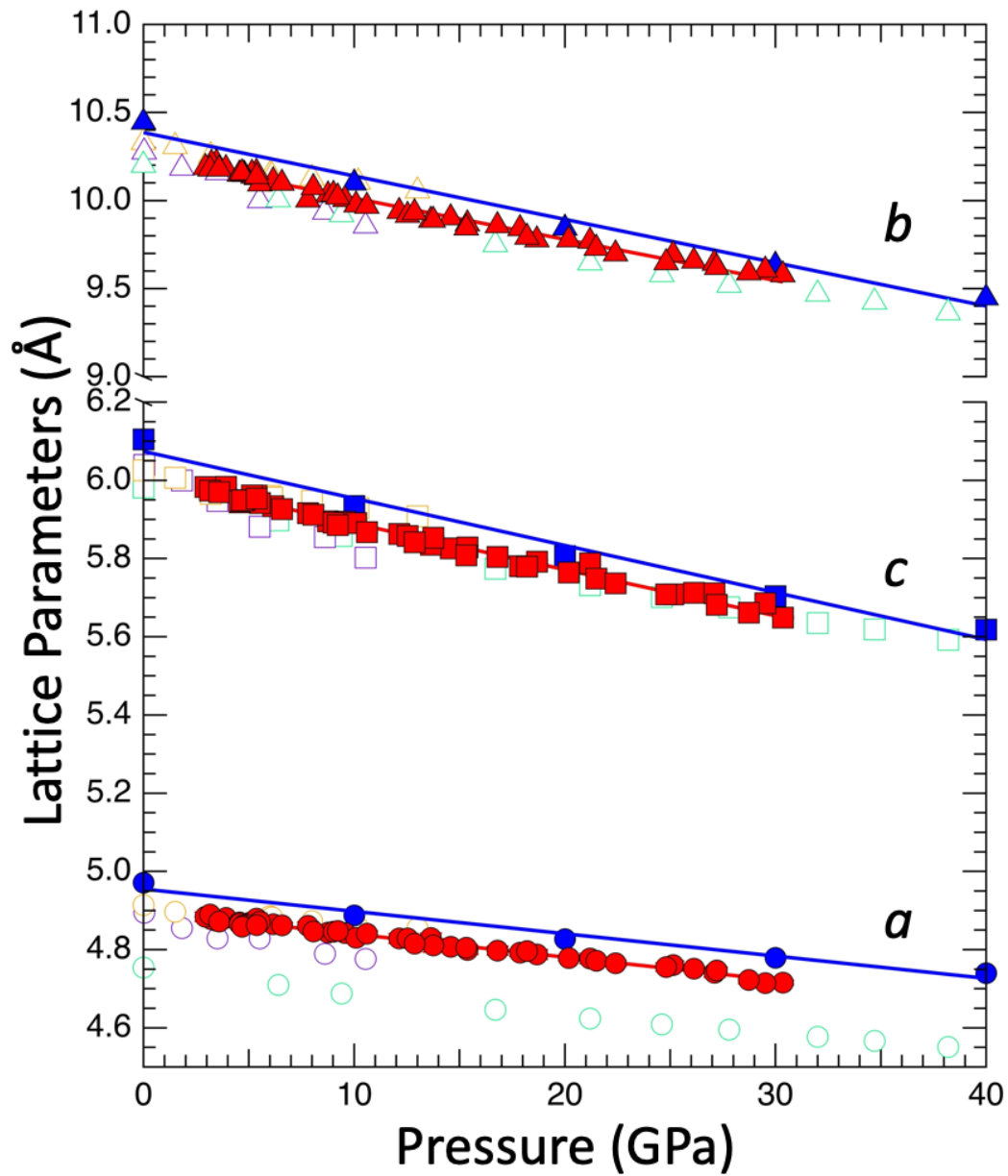

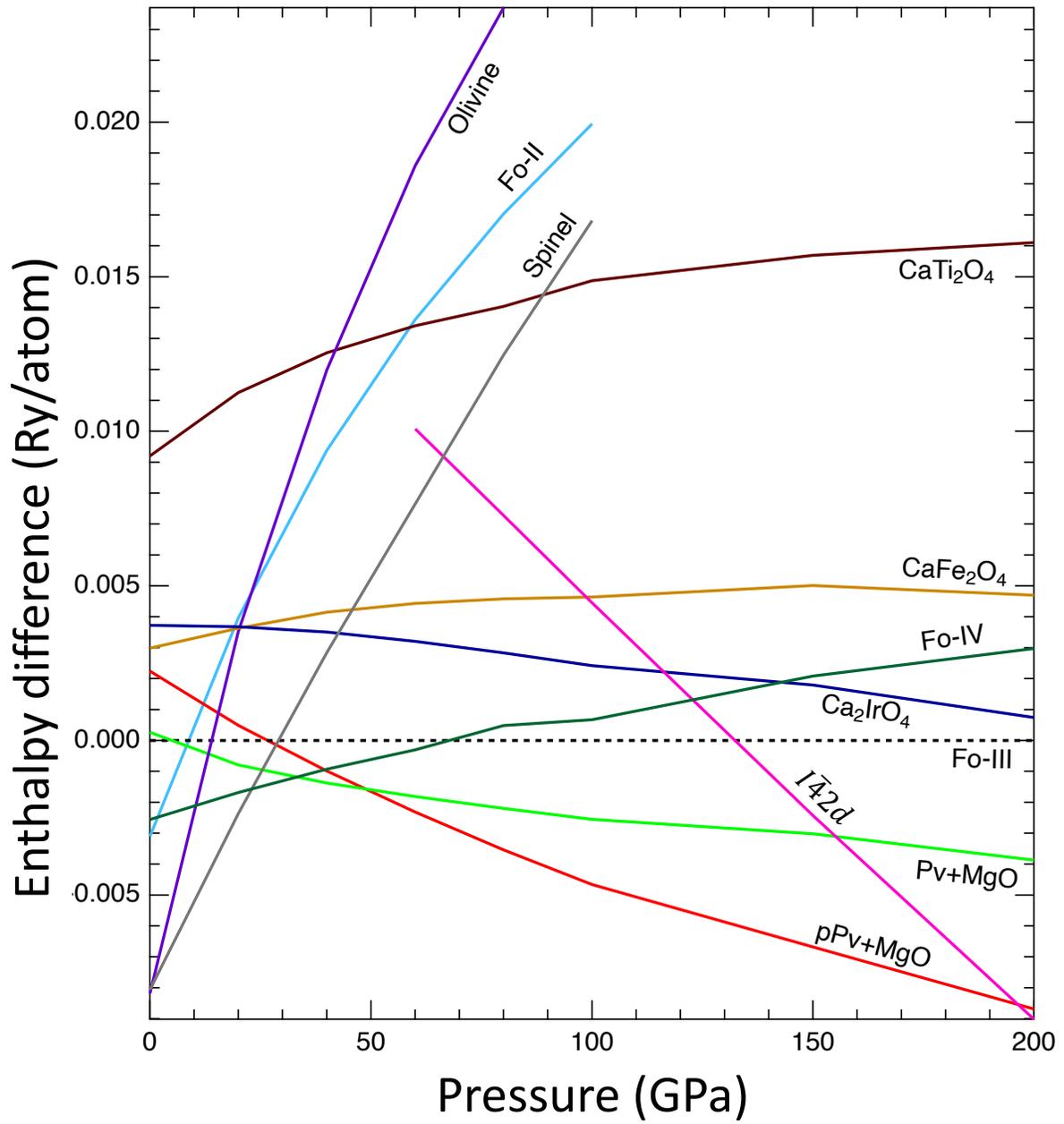

Figure 4. Enthalpy difference of the different phases of $Mg_2GeO_4$ with respect to the Forsterite-III type phase at 0 K.

Figure 5. Comparison of the observed XRD pattern at 61 GPa with the simulated patterns of the computed Fo-II, Fo-III and *Pmma* CaTi$_2$O$_4$-type Mg$_2$GeO$_4$ structures at 60 GPa.

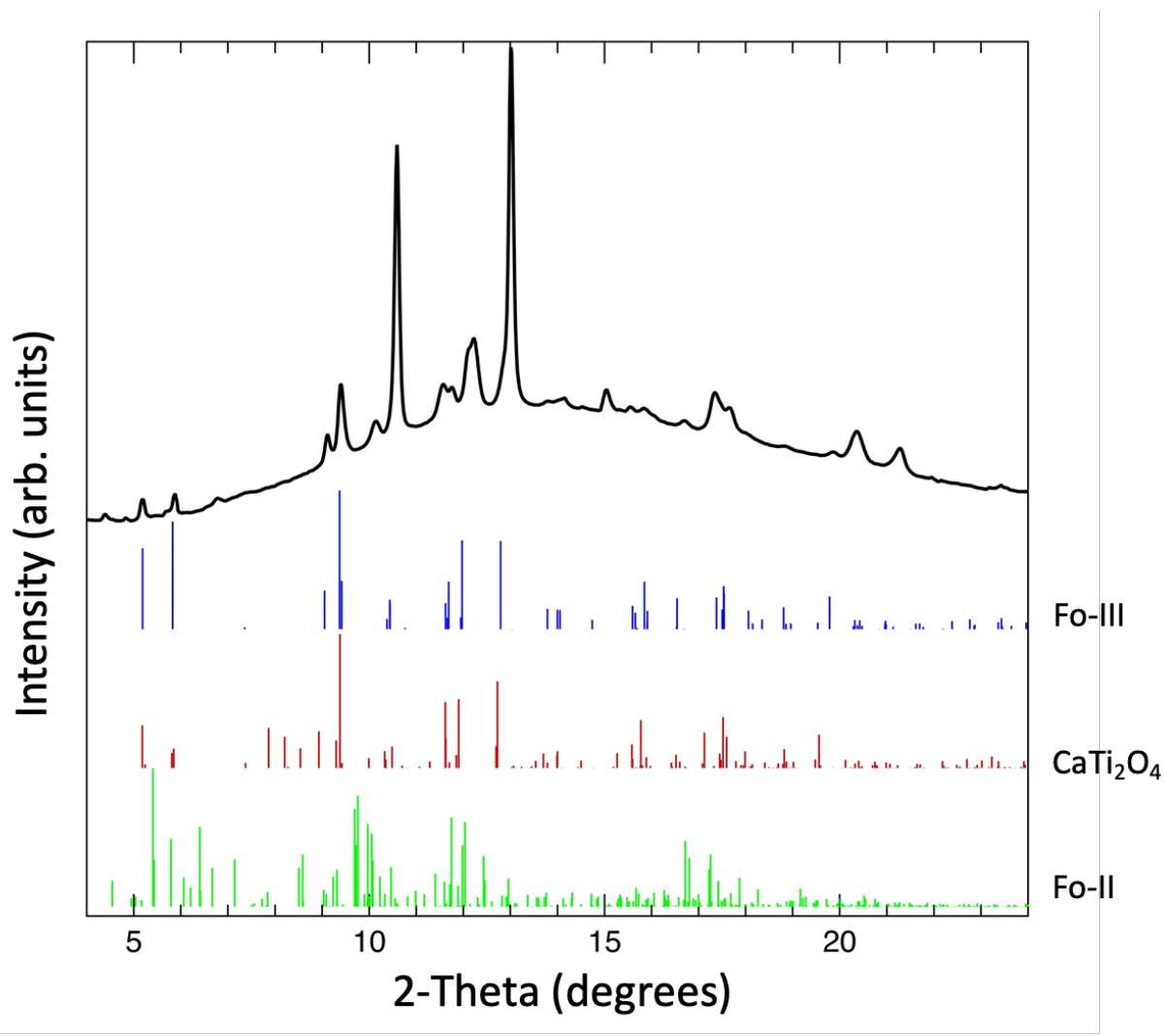

Figure 6. Crystal structure of forsterite- and forsterite-III type $Mg_2GeO_4$. Mg1 and Mg2 indicate the two inequivalent magnesium sites.

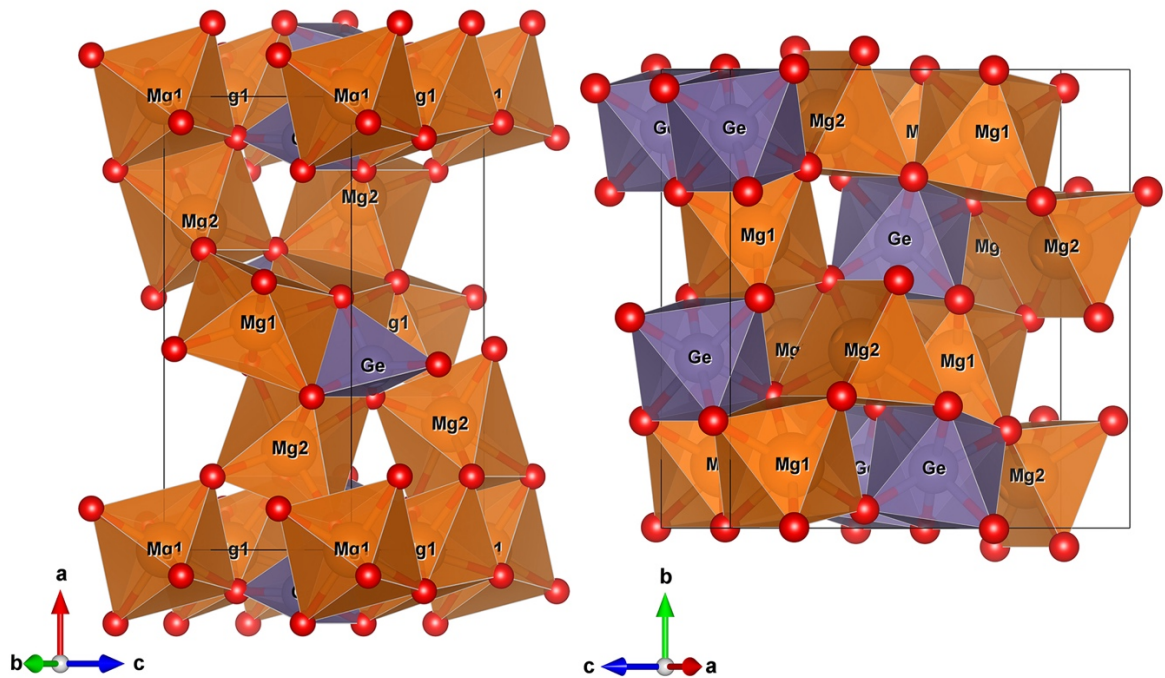

Figure 7. Le Bail refinement of XRD pattern of $Mg_2GeO_4$ at 74 GPa and 300 K. Colors have the same meaning as figure 2.

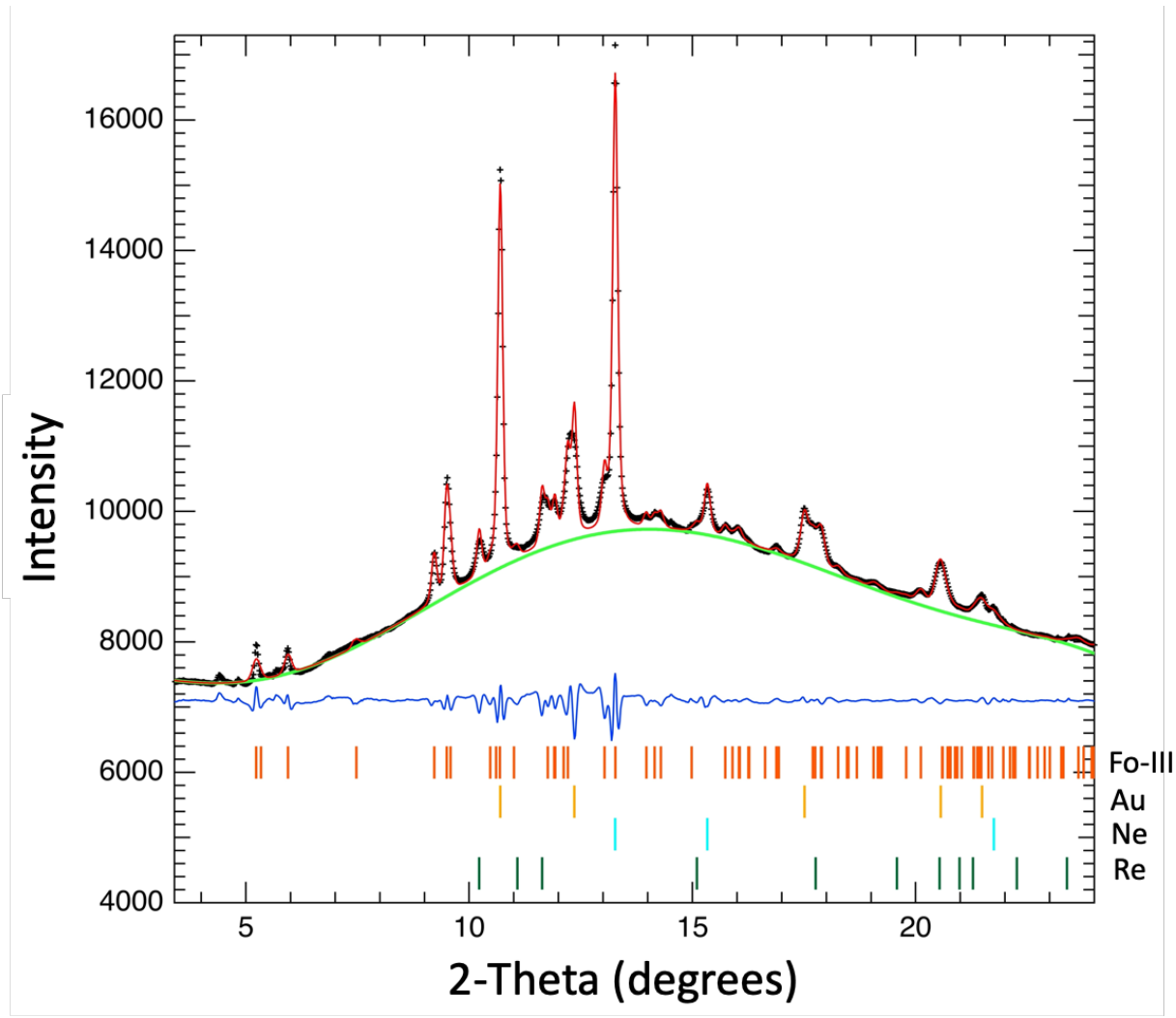

Figure 8. Lattice parameters of Fo-III type $Mg_2GeO_4$ versus pressure. Solid orange and blue indicate our experimental and theoretical data respectively. Open green symbols show the $Mg_2SiO_4$ data (Finkelstein et al., 2014).

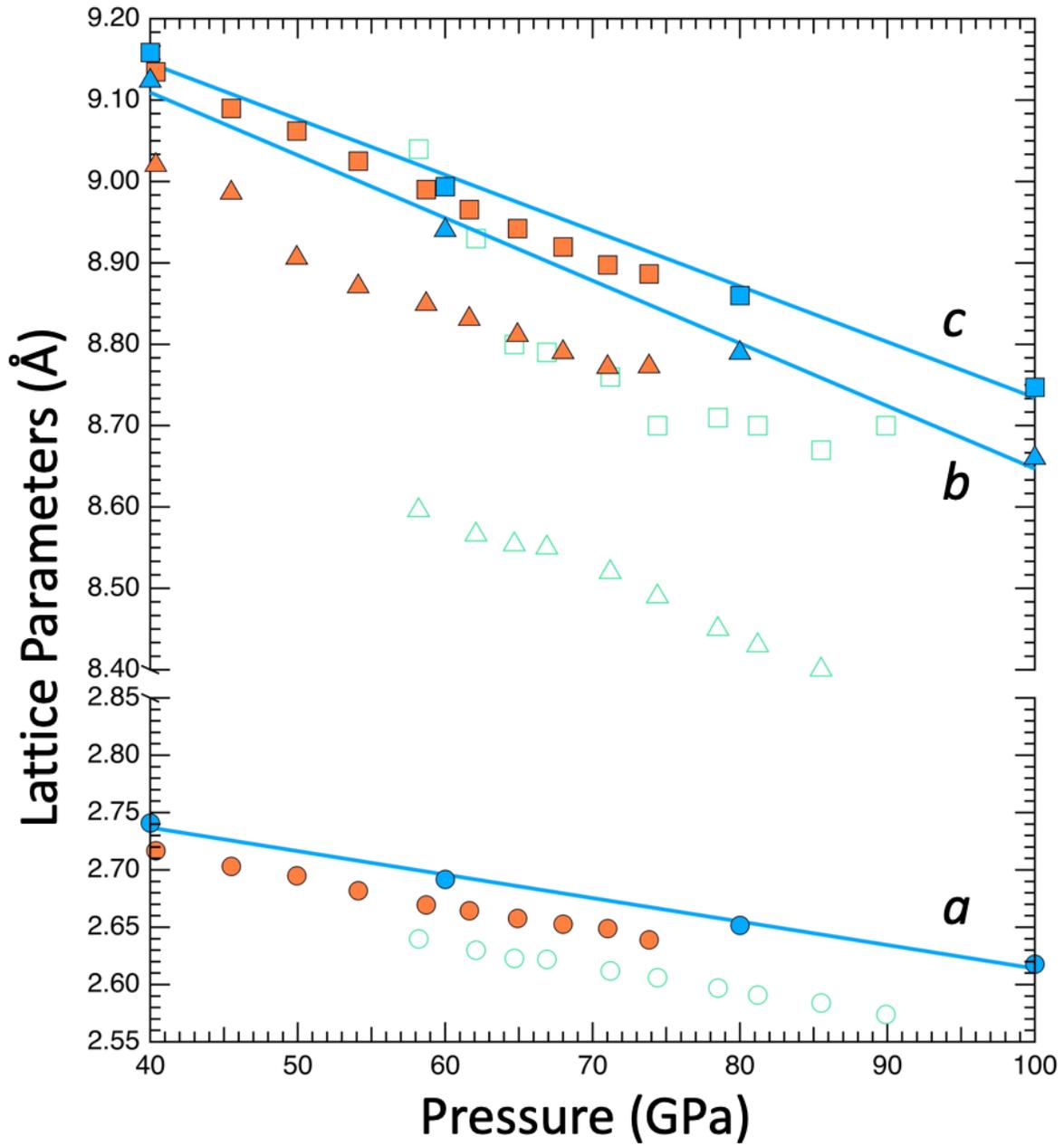

Figure 9. Variation in unit cell volume as a function of pressure. Solid circles (red: Forsterite, orange: Fo-III) and triangles (dark blue: Fo, light blue: FoIII) represent experimental and theoretical data from this study. Solid lines are 3$^{rd}$ order BM fits to the data. The two shades of green show the silicate data for Fo and Fo-III respectively (Finkelstein et al., 2014). Other colors have the same meaning as the figure 3.

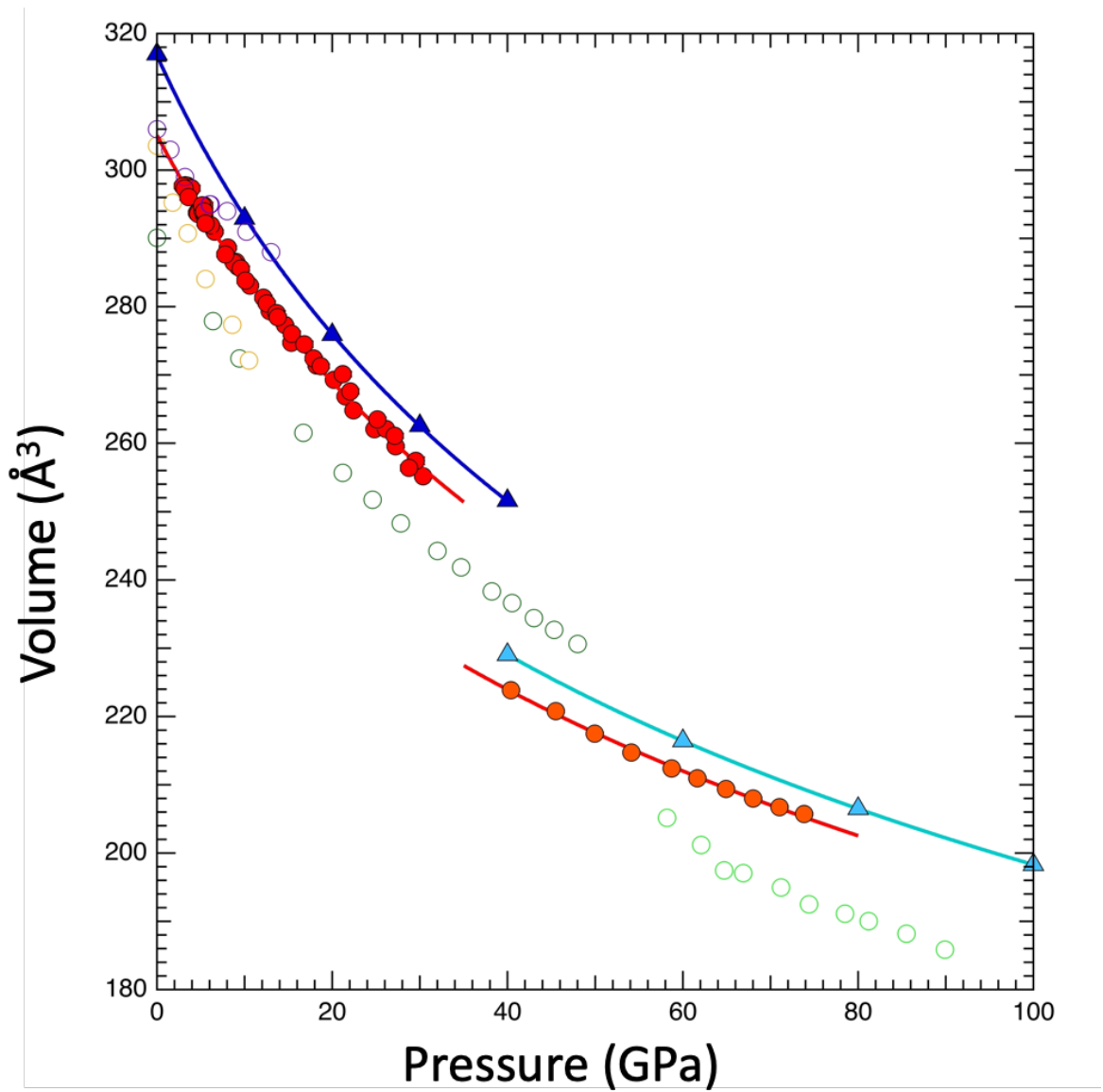

Figure 10. Le Bail refinement of the X-ray diffraction pattern of $Mg_2GeO_4$ after laser heating to 2460 K and then quenching to room temperature at 65 GPa. Colors have the same meaning as figure 2.

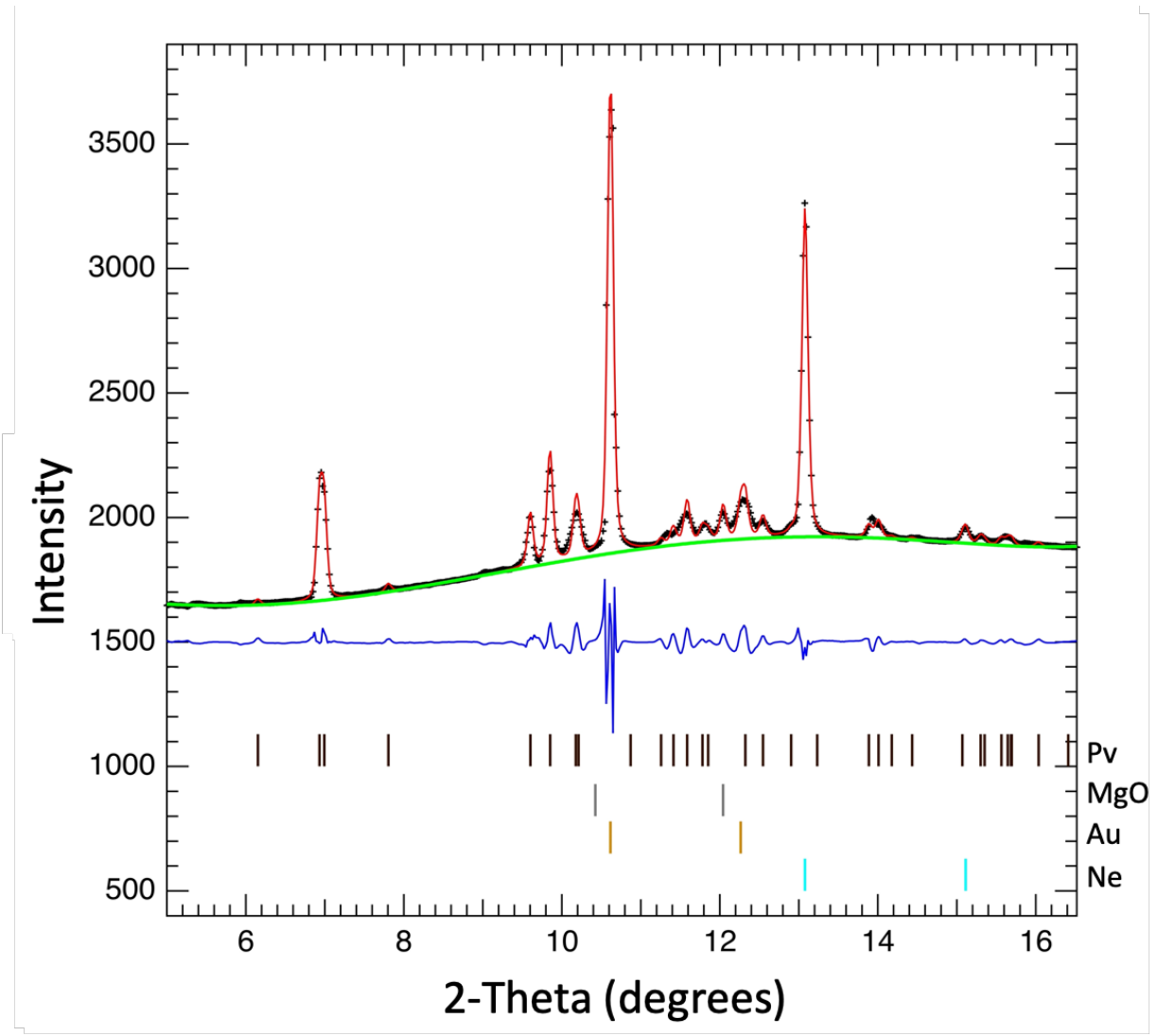

Figure 11. Le Bail refinement of the diffraction pattern of $Mg_2GeO_4$ at 110 GPa and 2280 K. Colors have the same meaning as figure 2.

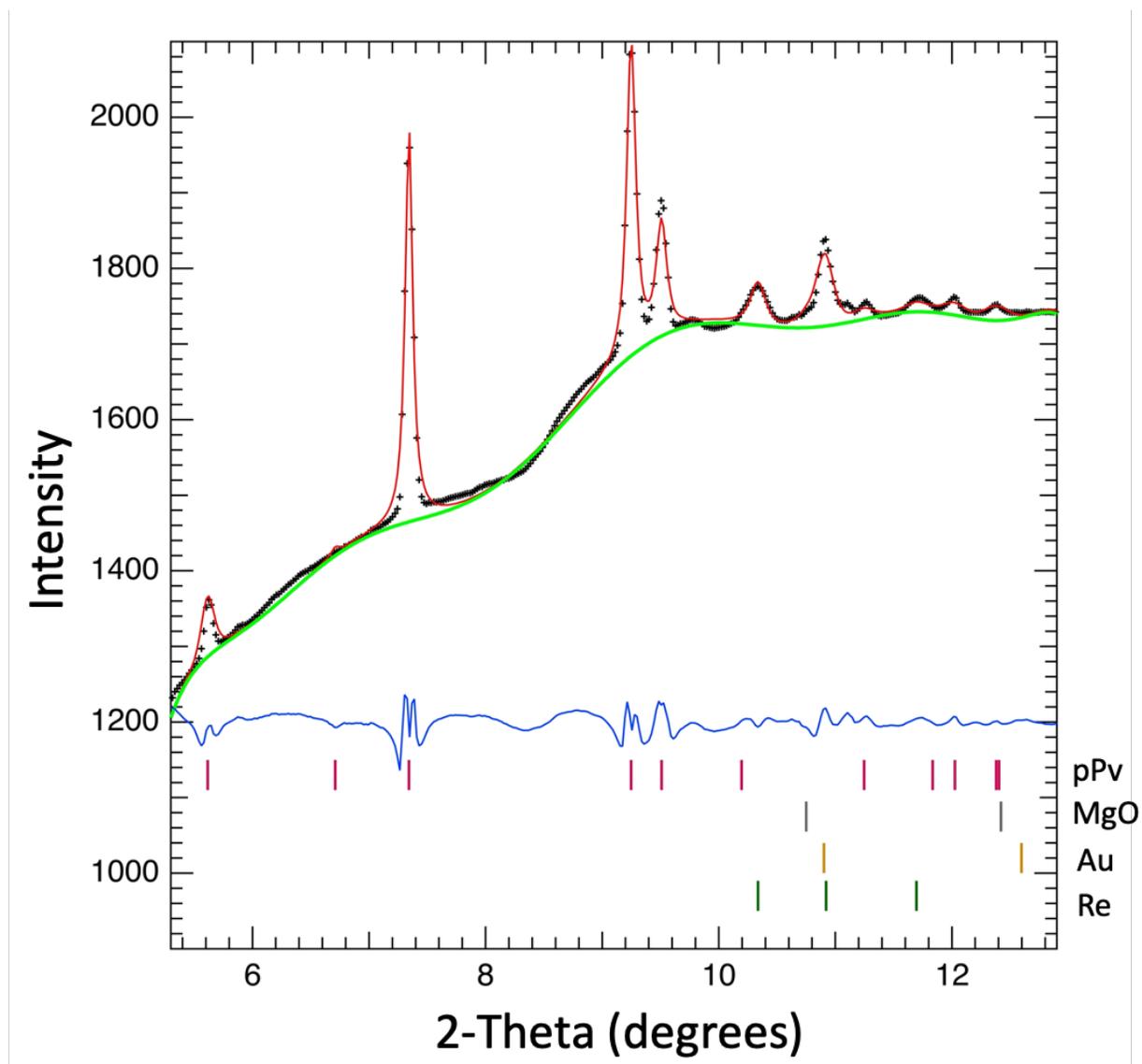

Table 1. Structural parameters of forsterite- and forsterite-III type $Mg_2GeO_4$.

| Phase | | Lattice Parameters | | | Atomic Positions | | | |
| --- | --- | --- | --- | --- | --- | --- | --- | --- |
| | | a (Å) | b (Å) | c (Å) | Atom | x | y | z |
| Forsterite-type | DFT (0 GPa) | 10.445 | 6.104 | 4.971 | Mg1 | 0 | 0 | 0 |
| | | | | | Mg2 | 0.77617 | 0.25 | 0.49341 |
| | | | | | Ge | 0.59424 | 0.25 | -0.06409 |
| | | | | | O1 | 0.59210 | 0.25 | 0.28914 |
| | | | | | O2 | -0.05998 | 0.25 | 0.73013 |
| | | | | | O3 | 0.66431 | 0.02121 | 0.76671 |
| | Exp (2.9 GPa) | 4.884 (7) | 10.188 (7) | 5.983 (4) | | | | |
| Fo-III type | DFT (60 GPa) | 2.692 | 8.940 | 8.994 | Mg1 | 0 | 0.86261 | 0.34586 |
| | | | | | Mg2 | 0 | 0.38687 | 0.67328 |
| | | | | | Ge | 0 | 0.87087 | -0.00852 |
| | | | | | O1 | 0 | 0.49826 | 0.42649 |
| | | | | | O2 | 0 | 0.76602 | 0.54234 |
| | | | | | O3 | 0 | 0.21248 | 0.30784 |
| | | | | | O4 | 0 | -0.04071 | 0.16471 |
| | Exp (61.6 GPa) | 2.664 (1) | 8.831 (7) | 8.966 (6) | | | | |

Table 2. Equation of state parameters for forsterite- and forsterite-III type $Mg_2GeO_4$ and $Mg_2SiO_4$.

| Phase | Mg₂GeO₄ | | | | Mg₂SiO₄ | | | |
|---|---|---|---|---|---|---|---|---|
| | $V_0$ (Å³) | $K_0$ (GPa) | $K_0'$ | Reference | $V_0$ (Å³) | $K_0$ (GPa) | $K_0'$ | Reference |
| Forsterite-type | 316.8 (3) | 112.2 (13) | 3.86 (5) | This study (DFT) | 290.1 (1) | 130.0 (9) | 4.12 (7) | Finkelstein et al., 2014 (Exp) |
| | 305.1 (3) | 124.6 (14) | 3.86 (fixed) | This study (Exp) | 289.17 | 128 (8) | 4 (fixed) | Andrault et al., 1995 (Exp) |
| | 303 | 70 (5) | - | Nagai et al., 1994 (Exp) | 289.3 (1) | 128.8 (5) | 4.2 (2) | Zhang., 1998 (Exp) |
| | 306 (4) | 166 (15) | 4 (fixed) | Petit et al., 1996 (Exp) | | | | |
| | 305.4 | 120 | - | Weidner and Hamaya., 1984 (Exp) | 290.14 (9) | 125 (2) | 4.0 (4) | Downs et al., 1996 (Exp) |
| Forsterite-III type | 271.8 (9) | 162.9 (5) | 4.19 (1) | This study (DFT) | 247.4 | 197 | 3.4 | Bouibes and Zaoui., 2020 (DFT) |
| | 263.5 (15) | 175 (7) | 4.19 (fixed) | This study (Exp) | 249.17 | 184.9 (8) | 4.11 (5) | Zhang et al., 2019 (DFT) |

<u>**Supplementary Material**</u>

**High-pressure Phase Transition of Olivine-type $Mg_2GeO_4$ to a Metastable Forsterite-III type Structure and their Equation of States.**

R. Valli Divya[1], Gulshan Kumar[1], R. E. Cohen[2], Sally J. Tracy[2], Yue Meng[3], Stella Chariton[4], Vitali B. Prakapenka[4], and Rajkrishna Dutta[1, *]

[1]Department of Earth Sciences, IIT Gandhinagar, Gujarat 382355, India.
[2]Earth and Planets Laboratory, Carnegie Institution for Science, Washington DC 20015, USA.
[3]HPCAT, Advanced Photon Source, Argonne National Laboratory, Argonne, IL 60439, USA.
[4]Center for Advanced Radiation Sources, University of Chicago, Chicago, IL 60637, USA.

Table S1. Observed ($d_{obs}$) and calculated ($d_{calc}$) $d$-spacings and their difference for olivine Mg$_2$GeO$_4$ at 14.6 GPa.

| h | k | l | $d_{obs}$ (Å) | $d_{calc}$ (Å) | $d_{obs} - d_{calc}$ |
|---|---|---|---|---|---|
| 2 | 0 | 0 | 4.95201 | 4.95116 | 0.00085 |
| 1 | 0 | 1 | 4.32317 | 4.32473 | -0.00156 |
| 2 | 1 | 0 | 3.77228 | 3.77279 | -0.00051 |
| 0 | 1 | 1 | 3.70813 | 3.70804 | 0.00009 |
| 2 | 1 | 1 | 2.96979 | 2.96796 | 0.00183 |
| 0 | 2 | 0 | 2.91275 | 2.91302 | -0.00027 |
| 3 | 0 | 1 | 2.72168 | 2.72112 | 0.00055 |
| 3 | 1 | 1 | 2.46399 | 2.46546 | -0.00147 |
| 1 | 2 | 1 | 2.41537 | 2.41605 | -0.00068 |
| 2 | 2 | 1 | 2.22637 | 2.22548 | 0.00088 |

Table S2. Lattice parameters of Mg$_2$GeO$_4$ olivine at different pressures.

| Pressure (GPa) | a (Å) | b (Å) | c (Å) | Volume (Å$^3$) |
|---|---|---|---|---|
| 5.2 | 4.872 (2) | 10.140 (4) | 5.952 (3) | 294.1 (2) |
| 6.2 | 4.865 (2) | 10.113 (6) | 5.934 (3) | 292.0 (2) |
| 8.8 | 4.844 (4) | 10.032 (6) | 5.895 (3) | 286.5 (2) |
| 14.6 | 4.807 (4) | 9.902 (6) | 5.826 (3) | 277.3 (2) |
| 20.2 | 4.779 (2) | 9.776 (6) | 5.764 (3) | 269.3 (2) |
| 26.1 | 4.752 (5) | 9.656 (6) | 5.713 (3) | 262.1 (2) |
| 30.4 | 4.716 (5) | 9.578 (6) | 5.649 (3) | 255.2 (2) |
| 3.5 | 4.874 (8) | 10.215 (17) | 5.981 (4) | 297.7 (4) |
| 3.2 | 4.878 (8) | 10.212 (17) | 5.978 (4) | 297.8 (4) |
| 2.9 | 4.884 (7) | 10.188 (11) | 5.983 (4) | 297.7 (4) |
| 3.9 | 4.881 (7) | 10.181 (12) | 5.984 (5) | 297.4 (4) |
| 3.1 | 4.890 (7) | 10.179 (11) | 5.974 (4) | 297.3 (4) |
| 3.6 | 4.872 (7) | 10.179 (14) | 5.970 (4) | 296.1 (4) |
| 5.4 | 4.867 (7) | 10.161 (14) | 5.961 (4) | 294.8 (4) |
| 5.1 | 4.871 (8) | 10.154 (17) | 5.962 (4) | 294.9 (4) |
| 5.4 | 4.879 (7) | 10.130 (14) | 5.947 (5) | 294.0 (4) |
| 5.5 | 4.871 (6) | 10.095 (10) | 5.942 (4) | 292.2 (3) |
| 7.8 | 4.861 (6) | 10.003 (10) | 5.916 (4) | 287.7 (3) |
| 9.6 | 4.843 (7) | 10.006 (13) | 5.894 (4) | 285.6 (3) |
| 10.1 | 4.831 (7) | 9.975 (10) | 5.891 (4) | 283.8 (4) |
| 12.1 | 4.829 (7) | 9.938 (12) | 5.863 (4) | 281.3 (3) |
| 12.5 | 4.829 (7) | 9.916 (12) | 5.858 (4) | 280.5 (3) |
| 13.6 | 4.825 (7) | 9.881 (12) | 5.853 (4) | 279.1 (3) |
| 13.8 | 4.811 (7) | 9.888 (17) | 5.854 (4) | 278.5 (4) |
| 15.4 | 4.799 (7) | 9.868 (11) | 5.829 (4) | 276.0 (4) |
| 16.8 | 4.797 (7) | 9.858 (11) | 5.805 (5) | 274.5 (4) |
| 17.9 | 4.786 (6) | 9.819 (9) | 5.797 (4) | 272.4 (3) |
| 18.7 | 4.778 (6) | 9.806 (9) | 5.791 (4) | 271.3 (3) |
| 21.2 | 4.778 (6) | 9.768 (13) | 5.788 (4) | 270.1 (4) |
| 22.1 | 4.794 (7) | 9.725 (10) | 5.740 (4) | 267.6 (4) |
| 25.1 | 4.770 (6) | 9.655 (9) | 5.722 (4) | 263.5 (3) |
| 27.1 | 4.741 (7) | 9.643 (12) | 5.711 (4) | 261.1 (3) |
| 29.5 | 4.714 (12) | 9.605 (9) | 5.686 (4) | 257.5 (5) |
| 28.7 | 4.723 (7) | 9.589 (9) | 5.662 (5) | 256.4 (4) |
| 4.8 | 4.866 (4) | 10.148 (6) | 5.950 (3) | 293.8 (2) |
| 4.6 | 4.870 (4) | 10.150 (7) | 5.944 (4) | 293.8 (3) |
| 4.6 | 4.866 (4) | 10.149 (6) | 5.947 (3) | 293.7 (3) |
| 4.7 | 4.862 (4) | 10.151 (7) | 5.950 (3) | 293.6 (2) |
| 4.7 | 4.859 (4) | 10.157 (7) | 5.949 (4) | 293.6 (3) |
| 5.4 | 4.863 (4) | 10.136 (7) | 5.954 (4) | 293.5 (3) |
| 6.6 | 4.862 (4) | 10.099 (7) | 5.926 (3) | 291.0 (2) |
| 8.1 | 4.847 (5) | 10.074 (7) | 5.912 (3) | 288.7 (2) |
| 9.0 | 4.846 (4) | 10.036 (8) | 5.892 (3) | 286.6 (2) |
| 9.2 | 4.848 (4) | 10.020 (7) | 5.886 (3) | 285.9 (2) |
| 10.6 | 4.841 (4) | 9.967 (7) | 5.868 (3) | 283.1 (2) |
| 12.9 | 4.815 (5) | 9.932 (7) | 5.842 (3) | 279.4 (2) |
| 15.3 | 4.805 (5) | 9.844 (7) | 5.808 (3) | 274.8 (2) |
| 18.2 | 4.801 (5) | 9.784 (9) | 5.778 (3) | 271.4 (2) |
| 21.5 | 4.772 (5) | 9.729 (7) | 5.749 (3) | 266.9 (2) |
| 22.4 | 4.762 (5) | 9.701 (8) | 5.733 (3) | 264.9 (2) |
| 24.8 | 4.765 (5) | 9.640 (7) | 5.705 (3) | 262.1 (2) |
| 27.2 | 4.741 (5) | 9.632 (7) | 5.684 (3) | 259.6 (2) |

Table S3. Observed ($d_{obs}$) and calculated ($d_{calc}$) d-spacings and their difference for Fo-III type Mg$_2$GeO$_4$ at 68.0 GPa.

| h | k | l | $d_{obs}$ (Å) | $d_{calc}$ (Å) | $d_{obs} - d_{calc}$ |
|---|---|---|---|---|---|
| 0 | 0 | 2 | 4.4664 | 4.45989 | 0.00651 |
| 0 | 2 | 1 | 3.94582 | 3.94249 | 0.00332 |
| 1 | 1 | 0 | 2.53975 | 2.53962 | 0.00014 |
| 0 | 2 | 3 | 2.46088 | 2.46268 | -0.0018 |
| 1 | 3 | 1 | 1.92031 | 1.92041 | -0.0001 |

Table S4. Lattice parameters of Fo-III type Mg$_2$GeO$_4$ at different pressures.

| Pressure (GPa) | a (Å) | b (Å) | c (Å) | Volume (Å$^3$) |
|---|---|---|---|---|
| 40.4 | 2.717 (2) | 9.020 (8) | 9.135 (6) | 223.9 (2) |
| 45.5 | 2.703 (2) | 8.986 (8) | 9.090 (6) | 220.8 (2) |
| 49.9 | 2.695 (2) | 8.906 (8) | 9.062 (6) | 217.5 (2) |
| 54.1 | 2.682 (2) | 8.871 (8) | 9.025 (6) | 214.7 (2) |
| 58.7 | 2.670 (1) | 8.849 (7) | 8.990 (6) | 212.4 (2) |
| 61.6 | 2.664 (1) | 8.831 (7) | 8.966 (6) | 211.0 (2) |
| 64.9 | 2.658 (1) | 8.811 (7) | 8.942 (6) | 209.4 (2) |
| 68.0 | 2.653 (1) | 8.790 (7) | 8.920 (6) | 208.0 (2) |
| 71.0 | 2.649 (1) | 8.772 (7) | 8.898 (6) | 206.7 (2) |
| 73.8 | 2.639 (1) | 8.773 (7) | 8.887 (6) | 205.7 (2) |